\documentclass[aps,twocolumn,preprintnumbers,superscriptaddress,
              showpacs,nofootinbib]{revtex4}
 \newcommand{\PRE}[1]{}
\usepackage{amsmath,amsfonts,epsfig,graphicx}
\usepackage{multirow}
\usepackage[latin1]{inputenc}
\usepackage{amssymb}
\usepackage{verbatim}
\usepackage{float}
\usepackage{slashed}

\newcommand{\MGMCatNLO}{MadGraph5\_aMC@NLO}

\newcommand{\pt}{p_\mathrm{T}}
\newcommand{\pythia}{{\sc Pythia}}

\newcommand{\delphes}{{\sc Delphes}}

\newcommand{\PZ}{\mathrm{Z}}
\newcommand{\PW}{\mathrm{W}}
\newcommand{\PH}{\mathrm{H}}
\newcommand{\Pb}{\mathrm{b}}

\begin{document}

\title{The physics case for an electron-muon collider}

\author{Meng Lu}
\affiliation{
Department of Physics and State Key Laboratory of Nuclear Physics and Technology, Peking University, Beijing, 100871, China}

\author{Andrew Michael Levin}
\affiliation{
Department of Physics and State Key Laboratory of Nuclear Physics and Technology, Peking University, Beijing, 100871, China}

\author{Congqiao Li}
\affiliation{
Department of Physics and State Key Laboratory of Nuclear Physics and Technology, Peking University, Beijing, 100871, China}

\author{Antonios Agapitos}
\affiliation{
Department of Physics and State Key Laboratory of Nuclear Physics and Technology, Peking University, Beijing, 100871, China}

\author{Qiang Li}
\affiliation{
Department of Physics and State Key Laboratory of Nuclear Physics and Technology, Peking University, Beijing, 100871, China}

\author{Fanqiang Meng}
\affiliation{
Department of Physics and State Key Laboratory of Nuclear Physics and Technology, Peking University, Beijing, 100871, China}

\author{Sitian Qian}
\affiliation{
Department of Physics and State Key Laboratory of Nuclear Physics and Technology, Peking University, Beijing, 100871, China}

\author{Jie Xiao}
\affiliation{
Department of Physics and State Key Laboratory of Nuclear Physics and Technology, Peking University, Beijing, 100871, China}

\author{Tianyi Yang}
\affiliation{
Department of Physics and State Key Laboratory of Nuclear Physics and Technology, Peking University, Beijing, 100871, China}

\begin{abstract}
An electron-muon collider with an asymmetric collision profile targeting multi-ab$^{-1}$ integrated luminosity is proposed. This novel collider, operating at collisions energies of e.g. 20--200\,GeV, 50--1000\,GeV and 100--3000\,GeV, would be able to probe charged lepton flavor violation and measure Higgs boson properties precisely. The collision of an electron and muon beam leads to less physics background compared with either an electron-electron or a muon-muon collider, since electron-muon interactions proceed mostly through higher order vector boson fusion and vector boson scattering processes. The asymmetric collision profile results in collision products that are boosted towards the electron beam side, which can be exploited to reduce beam-induced background from the muon beam to a large extent. With this in mind, one can imagine a lepton collider complex, starting from colliding order 10 GeV electron and muon beams for the first time in history and to probe charged lepton flavor violation, then to be upgraded to a collider with 50-100 GeV electron and 1-3 TeV muon beams to measure Higgs properties and search for new physics, and finally to be transformed to a TeV scale muon muon collider. The cost should vary from order 100 millions to a few billion dollars, corresponding to different stages, which make the funding situation more practical.\\  \texttt{meng.lu@cern.ch, jie.xiao@cern.ch, qliphy0@pku.edu.cn, andrew.michael.levin@cern.ch}

\end{abstract}


\maketitle

The discovery~\cite{plb:2012gu,plb:2012gk} and property measurements~\cite{higprop} of the Higgs boson are a triumph of the Standard Model (SM) of particle physics and also the Large Hadron Collider (LHC). In the next 10-20 years,the LHC and the High-Luminosity LHC (HL-LHC) will be further exploring the SM and searching for physics beyond that. The 2020 update of the European Strategy for Particle Physics~\cite{UES} proposes a vision for both the near- and the long-term future of the field, highlighting the need to pursue, as the highest-priority facility after the LHC, an electron-electron collider acting as a Higgs factory, such as the proposed International Linear Collider (ILC), the Compact Linear Collider (CLIC), the Future Circular Collider (FCC), or the Circular Electron Positron Collider (CEPC). These future colliders may, at various stages in their lifecycles, also serve as $\PZ$ or top quark factories or operate at the TeV scale for the purpose of new physics searches. The FCC and CEPC projects also aim for an upgrade to an order  100\,TeV hadron collider. Other colliders that have been proposed with similar objectives are the High Energy LHC (HE-LHC)~\cite{HELHC}, a Large Hadron electron Collider (LHeC)~\cite{LHeC}, and a muon collider~\cite{MuC}.

Recently there has been a large amount of interest~\cite{Daniel20} in a muon-muon collider with  center-of-mass energies in the multi-TeV range, which has some of the advantages of both hadron-hadron and electron-electron colliders~\cite{Mario16,Antonio20,Dario18}. Because muon beams emit much less synchrotron radiation than electron beams, muons can be accelerated in a circular collider to much higher energies. On the other hand, because the proton is a composite particle, muon-muon collisions are cleaner than proton-proton collisions and also can lead to higher effective center-of-mass energies. However, due to the short lifetime of the muon, the beam-induced background (BIB) from muon decays needs to be examined and reduced properly. Based on a realistic simulation at $\sqrt{s}=1.5$\,TeV with BIB included, Ref.~\cite{Nazar20} found that the coupling between the Higgs boson and the b-quark can be measured at percent level with order ab$^{-1}$ of collected data.

Here we propose an electron-muon collider with an asymmetric collision profile of, e.g., 20--200\,GeV, 50--1000\,GeV and 100--3000\,GeV for the electron-muon beam energy, respectively, corresponding to the center-of-mass energy as 126.5\,GeV, 447.2\,GeV, and 1095.4\,GeV. In history, CERN and Brookhaven scattered muons in the 1960s, using accelerator-produced muons at GeV scale on electrons in targets, and measured the elastic scattering cross section~\cite{muone1963}. In the late 1990s the scattering of muons off polarized electrons was used by the SMC collaboration at CERN as a polarimeter for high-energy muon beams~\cite{muone1990}. Recently, a new experiment, MUonE~\cite{muone}, has been proposed, through elastically scattering of high-energy muons at around 150\,GeV on atomic electrons, to determine the leading hadronic contribution to the muon $g$-2. A head-on electron-muon collider, was first proposed in the 1990s~\cite{hou96,choi97,barger97} mainly to probe charged lepton flavor violation (CLFV)~\cite{fabio20}, which, however, has not been followed up much by the community. Here we propose for the first time, an electron-muon collider with more detailed configurations to run at both low and high energy, with broader physics goals to cover both CLFV and Higgs studies.

We now elaborate on the physics case for an electron-muon collider. The asymmetric collision profile proposed above is motivated by the fact that muons can be accelerated to higher energies much more easily than electrons. An asymmetric electron-muon collider operating at the energies proposed above can serve as an intermediate step between an electron-electron collider and a muon-muon collider. The advantages of such a novel collider are that it enables a direct study of charged lepton flavor violation in a nearly background-free environment, in contrast with other lepton colliders. The collision of an electron and a muon beam leads to less physics background compared with either an electron-electron  or  a  muon-muon  beam because the  background physics  processes are mostly  higher order vector boson fusion or scattering processes. Moreover, the asymmetric beam profile tends to create collision products that are boosted towards the electron beam side, which can be exploited to reduce muon BIB from muon beam upstream to a large extent, for example, by adding a shielding nozzle in the muon side with a large cone size without much loss on acceptance.

\begin{figure}[htbp]
  \begin{center}
  \includegraphics[width=0.4\textwidth]{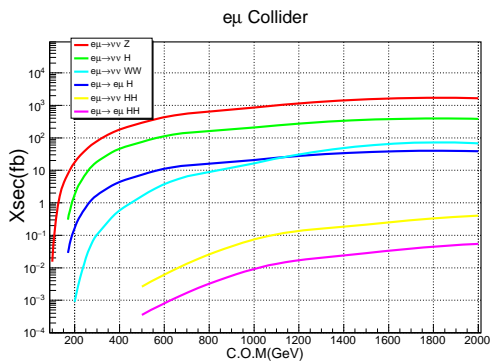}
    \caption{Cross section dependence on center-of-mass energy of six dominant physics processes at an electron-muon collider.}
    \label{fig:emucollider}
  \end{center}
\end{figure}

The cross section dependence on the center-of-mass energy for six of the dominant SM processes at an electron-muon collider is shown in Fig.~\ref{fig:emucollider}. The two leading processes in Fig.~\ref{fig:emucollider} are the vector boson fusion or scattering processes $e^-\mu^+\rightarrow \nu_e\tilde{\nu}_\mu \PZ$ and $e^-\mu^+\rightarrow \nu_e\tilde{\nu}_\mu \PH$. The cross section of $e^-\mu^+\rightarrow \nu_e\tilde{\nu}_\mu \PZ$ is 291\,fb (863\,fb) at $\sqrt{s}=0.5$\,TeV ($1.0$\,TeV) and the cross section of $e^-\mu^+\rightarrow \nu_e\tilde{\nu}_\mu \PH$ is 75\,fb (209\,fb) at $\sqrt{s}=0.5$\,TeV ($1.0$\,TeV).

\begin{figure}[htbp]
  \begin{center}
  \includegraphics[width=0.4\textwidth]{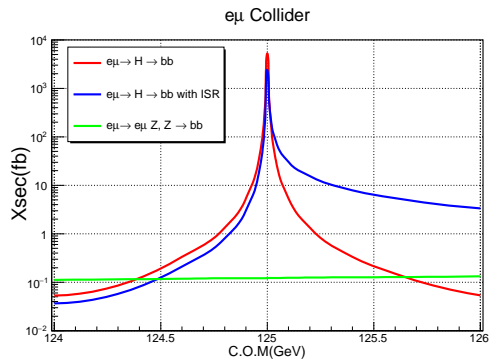}
    \caption{Energy scan near $\sqrt{s}=125$\,GeV for Higgs CLFV production $e\mu\rightarrow \PH \rightarrow \Pb\bar{\Pb}$ and the dominant background $e\mu\rightarrow \nu\nu \PZ \rightarrow \nu\nu \Pb\bar{\Pb}$ at an electron-muon collider. The CLFV Yukawa couplings $|Y_{e\mu}|$ is set by default to match to the current limit on Br($\PH\rightarrow e\mu$) from ATLAS experiment~\cite{ATLAS19}.}
    \label{fig:emulfv}
  \end{center}
\end{figure}

An electron-muon collider could operate at multiple energies to search for CLFV in $e\mu\rightarrow \PZ$ and $\PH$, or $e\mu\rightarrow e e$, etc. Here we take Higgs CLFV as an example (some other examples are discussed in~\cite{fabio20}), which is interesting because its observation may provide insight into some fundamental questions in nature, e.g., whether there is a secondary mechanism for the electroweak symmetry breaking. $\PH\rightarrow e\mu$ has already been studied at the LHC~\cite{CMS16, ATLAS19}, and the most stringent limit so far, from ATLAS, is Br($\PH\rightarrow e\mu$) $<6.2\times 10^{-5}$~\cite{ATLAS19}. The CEPC projected limit is Br($\PH\rightarrow e\mu$) $<1.2\times 10^{-5}$~\cite{Qin17}. In this paper, we implement the Higgs CLFV model~\cite{CMS16} in {\MGMCatNLO}~\cite{Alwall:2014hca}, with the CLFV Yukawa couplings $|Y_{e\mu}|$ set by default to match the current limit on Br($\PH\rightarrow e\mu$) from Ref.~\cite{ATLAS19}. In Fig.~\ref{fig:emulfv}, we show an energy scan near $\sqrt{s}=125$\,GeV (with e.g., 20--200\,GeV for the electron-muon beam energy) for Higgs CLFV production $e\mu\rightarrow \PH \rightarrow \Pb\bar{\Pb}$ with and without initial state radiation (ISR) effects~\cite{isr17}, and the dominant background $e\mu\rightarrow \nu_e\tilde{\nu}_\mu \PZ \rightarrow \nu_e\tilde{\nu}_\mu \Pb\bar{\Pb}$ at an electron-muon collider. The signal peak cross section is around 5.3\,pb, while the background cross section is around 0.1\,fb. Based on a simple estimation with the signal and background yields, with 1\,fb$^{-1}$ of data at around $\sqrt{s}=125$\,GeV, an electron-muon collider should already be able to obtain a 100 times better limit on Higgs CLFV than the current best result from ATLAS. As a caveat, the actual Higgs production rate will depend on the convolution of the muon beam energy profile with the Higgs production cross section, not just the peak cross section, as discussed in Ref.~\cite{Mario16}.

\begin{figure}[htbp]
  \begin{center}
  \includegraphics[width=0.4\textwidth]{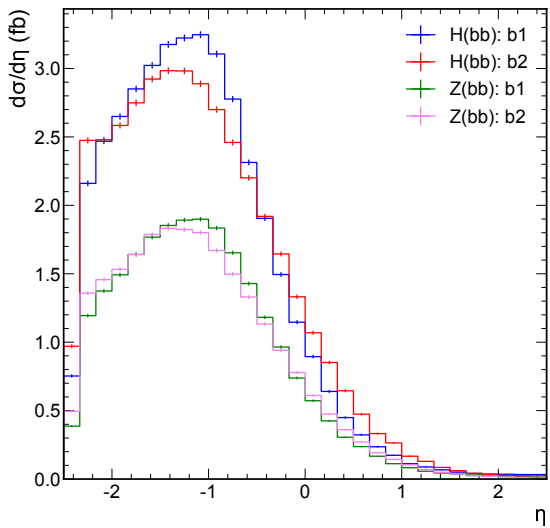}
    \caption{The $\eta$ distributions of the leading and subleading b-jet from Higgs and Z boson decay. The electron-beam (100\,GeV) and muon-beam (3000\,GeV) come from the negative and positive $\eta$ partition. The detector simulation is implemented through \delphes~ using the muon detector configuration. The Valencia algorithm~\cite{BORONAT} is used for jet reconstruction with the jet cone size as 0.5. The b-jets are required with $\pt>20$\,GeV and $|\eta| < 2.5$, with the b-tagging efficiency 70\%. The $\eta$ distributions indicate that jets are boosted towards the electron-beam side.}
    \label{fig:etabjets}
  \end{center}
\end{figure}

An electron-muon collider can also serve as a Higgs factory, with asymmetric collision profile of, e.g., 50--1000\,GeV and 100--3000\,GeV for the electron-muon beam energy, respectively, corresponding to the center-of-mass energy as 447.2\,GeV and 1095.4\,GeV. We would like to highlight two advantages of an electron-muon collider: (1) as there is no CLFV in the SM, one gets less physics backgrounds compared with the case of other same-flavor leptonic colliders. The dominant background to $e^-\mu^+\rightarrow \nu_e\tilde{\nu}_\mu \PH$ with $\PH\rightarrow \Pb\bar{\Pb}$ is $e^-\mu^+\rightarrow \nu_e\tilde{\nu}_\mu \PZ$ with $\PZ\rightarrow \Pb\bar{\Pb}$ which can be further reduced effectively by imposing selections on the invariant mass of di-b jets; (2) the asymmetric beam energies tends to have collision products boosted towards the electron beam side, as shown in Fig.~\ref{fig:etabjets}, which can be exploited to reduce muon BIB to a large extent, for example, by adding a shielding nozzle in the muon side with large cone size without much loss on acceptance. 

The cross section of $e^-\mu^+\rightarrow \nu_e\tilde{\nu}_\mu \PH$ with $\PH\rightarrow \Pb\bar{\Pb}$ depends on:
\begin{eqnarray}\label{equ1}
\sigma=\sigma(\nu\nu\PH) \cdot BR(\PH \rightarrow \Pb\bar{\Pb})=\frac{g^2_{\PH \PW \PW}g^2_{\PH \Pb \Pb}}{\Gamma_{\PH}},
\end{eqnarray} 
where $g_{\PH \PW \PW}$ is the coupling of the Higgs boson to the W boson, $g_{\PH \Pb \Pb}$ is the coupling of the Higgs boson to the b quark, and $\Gamma_{\PH}$ is the Higgs boson width. From the formula above, we can further obtain the uncertainty of $g_{\PH \Pb \Pb}$ as
\begin{eqnarray}\label{equ2}
\frac{\Delta g_{\PH \Pb \Pb}}{g_{\PH \Pb \Pb}}=\frac{1}{2}\sqrt{\left(\frac{\Delta \sigma}{\sigma}\right)^{2}+\left(\frac{\Delta \frac{g^2_{\PH \PW \PW}}{\Gamma_{\PH}}}{\frac{g^2_{\PH \PW \PW}}{\Gamma_{\PH}}}\right)^{2}}.
\end{eqnarray}

To roughly estimate the precision of a $g_{\PH \Pb \Pb}$ measurement with an electron-muon collider, we assume that ${g^2_{\PH \PW \PW}}/{\Gamma_{\PH}}$ can be measured with the same precision as in CLIC~\cite{CLIChiggs}, with a 3\% uncertainty conservatively. The cross section uncertainty can be approximated as
\begin{eqnarray}\label{equ3}
\frac{\Delta \sigma}{\sigma} \simeq \frac{\sqrt{N+B}}{N},
\end{eqnarray}
where $N$ and $B$ are the number of $\PH \rightarrow \Pb\bar{\Pb}$ and background events, respectively.

\begin{figure}[htbp]
  \begin{center}
  \includegraphics[width=0.4\textwidth]{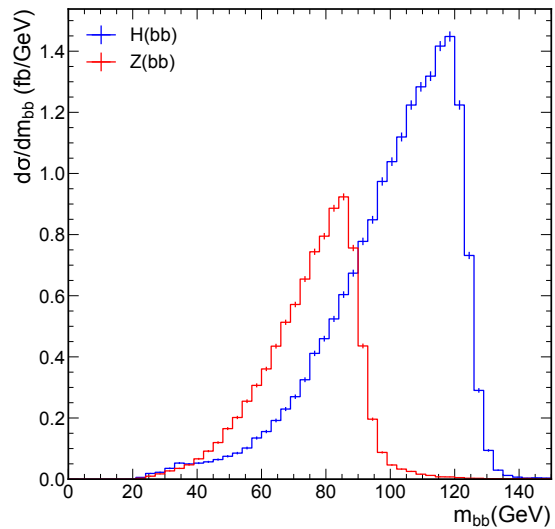}
    \caption{The invariant mass distribution of two b-tagged jets for the signal $\PH \rightarrow \Pb\bar{\Pb}$ and background $\PZ \rightarrow \Pb\bar{\Pb}$ process, after requiring $\pt^{\Pb}>40$\,GeV and $-2.5<\eta^{\Pb}<1$, at a 100--3000\,GeV electron-muon collider.}
    \label{fig:higgsmass}
  \end{center}
\end{figure}

 $\PH \rightarrow \Pb\bar{\Pb}$ and $\PZ \rightarrow \Pb\bar{\Pb}$ samples are generated using {\MGMCatNLO}~\cite{Alwall:2014hca} and interfaced with \pythia~\cite{Sjostrand:2003wg} for parton showering and hadronization. A fast simulation is performed using \delphes~version 3~\cite{deFavereau:2013fsa} with the muon collider configuration. We require that the leading and sub-leading b-tagged jets satisfy $\pt>40$\,GeV and $-2.5<\eta<1$ (corresponding to a 40.4$^{\circ}$ shielding nozzle on the muon beam side). Fig.~\ref{fig:higgsmass} shows the invariant mass distribution of the two b-tagged jets after this selection.

We require that $100 < m_{\rm{bb}} < 150\,\rm{GeV}$ to suppress the background and then evaluate the uncertainty on the cross section based on equation~(\ref{equ3}). Taking into account the ${g^2_{\PH \PW \PW}}\big/{\Gamma_{\PH}}$ uncertainty, we estimate the precision of $g_{\PH \Pb \Pb}$ in different collision schemes using the number of the signal and background events after the selection. The results are listed in Table~\ref{tab:coupling_results}, showing that the measured precision of $g_{\PH \Pb \Pb}$ at an electron-muon collider is at the few percent level with order ab$^{-1}$ of data and is dominated by the uncertainty on ${g^2_{\PH \PW \PW}}\big/{\Gamma_{\PH}}$.

\begin{table}[htb]
\centering
  \caption{Summary of collider parameters and the corresponding uncertainties on the Higgs boson coupling to b-quarks. $\sqrt{s}=447.2$  $(1095.3)$\,GeV corresponds to a 50 (100)\,GeV electron beam and a 1 (3)\,TeV muon beam.}
  \begin{tabular}{ccccc}
  \hline
  \hline
  ${\cal L}_{\mathrm{int}}$ $[ab^{-1}]$ & $\sqrt{s}$ [GeV] & $\frac{\Delta \sigma}{\sigma}$ [\%]& $\frac{\Delta \frac{g^2_{\PH \PW \PW}}{\Gamma_{\PH}}}{\frac{g^2_{\PH \PW \PW}}{\Gamma_{\PH}}}$ [\%]& $\frac{\Delta g_{\PH \Pb \Pb}}{g_{\PH \Pb \Pb}}$ [\%] \\
  \hline
  0.5 & \begin{tabular}{@{}c@{}}447.2 \\ 1095.4 \end{tabular} & \begin{tabular}{@{}c@{}}2.5 \\ 1.4 \end{tabular} & 3 & \begin{tabular}{@{}c@{}}2.0 \\ 1.7 \end{tabular} \\
  1.5 & \begin{tabular}{@{}c@{}}447.2 \\ 1095.4 \end{tabular} & \begin{tabular}{@{}c@{}}1.4 \\ 0.8 \end{tabular} & 3 & \begin{tabular}{@{}c@{}}1.7 \\ 1.6 \end{tabular} \\
  2.0 & \begin{tabular}{@{}c@{}}447.2 \\ 1095.4 \end{tabular} & \begin{tabular}{@{}c@{}}1.2 \\ 0.7 \end{tabular} & 3 & \begin{tabular}{@{}c@{}}1.6 \\ 1.6 \end{tabular} \\
  \hline
  \hline
  \end{tabular}
  \label{tab:coupling_results}
\end{table}

The physical infrastructure needed to generate electron and muon beams with energies below 100\,GeV would be either a small linear or circular accelerator. A muon beam at TeV scale can be achieved in a kilometer-sized storage ring. A set-up that can generate both electron and positron beams, or both muon and anti-muon beams, or polarized beams, should be considered to further enrich the physics outcome. Such a lepton collider complex, can start from colliding order 10 GeV electron and muon beams for the first time in history and to probe CLFV, then to be upgraded to a collider with 50-100 GeV electron and 1-3 TeV muon beams to measure Higgs properties and probe new physics beyond the SM including CLFV further, and finally to be transformed to a TeV scale muon muon collider. The cost should vary from order 100 millions to a few billion dollars~\cite{shiltsev17}, corresponding to different stages, which make the funding situation more practical. Note such a lepton collider complex can also serve as a muon source at various energy scales for general science research.

In summary, we have proposed a novel collider, with colliding electron and muon beams with asymmetric energies, and shown it has great potential to probe charged lepton flavor violation and to measure Higgs properties precisely. An electron-muon collider has less physics backgrounds than either an electron-electron collider or a muon-muon collider. The asymmetric beam energies lead to collision products that are boosted towards the electron beam side, which can be exploited to reduce muon BIB to a large extent. Such a collider can operate at center-of-mass energies up to a few TeV, and can serve as an intermediate step towards a future higher energy physics facility. We urge the community to consider this new option seriously. 

\pagebreak
\newpage
\acknowledgements
This work is supported in part by the National Natural Science Foundation of China under Grant No. 12075004, by MOST under grant No. 2018YFA0403900.


\begin{thebibliography}{10}

\bibitem{plb:2012gu} 
  S.~Chatrchyan {et al.}  [CMS Collaboration],
  Phys.\ Lett.\ B {\bf 716}, 30 (2012)
  [arXiv:1207.7235 [hep-ex]].

\bibitem{plb:2012gk} 
  G.~Aad {et al.}  [ATLAS Collaboration],
  Phys.\ Lett.\ B {\bf 716}, 1 (2012)
  [arXiv:1207.7214 [hep-ex]]. 

\bibitem{higprop} S. Chatrchyan, et al., (The CMS Collaboration), JHEP, {\bf 06}: 081 (2013); G. Aad, et al., (The ATLAS Collaboration), Phys. Lett. B, {\bf 726}: 88 (2013); G. Aad, et al., (The ATLAS Collaboration), Phys. Lett. B, {\bf 726}: 120 (2013); V. Khachatryan, et al., (The CMS Collaboration), Eur. Phys. J. C, {\bf 75}: 212 (2015); V. Khachatryan, et al., (The CMS Collaboration), Phys. Rev. D, {\bf 92}: 012004 (2015); G. Aad, et al., (The ATLAS Collaboration and CMS Collaboration), Phys. Rev. Lett., {\bf114}: 191803 (2015).
 
\bibitem{UES} https://europeanstrategy.cern/

\bibitem{LHeC} J. L. Abelleira Fernandez, et al., J. Phys. G, {\bf 39}, 075001(2012).

\bibitem{HELHC} X. Cid Vidal, et al., CERN-LPCC-2018-05.

\bibitem{MuC} Jean Pierre Delahaye, et al., arXiv:1901.06150.

\bibitem{Daniel20} Daniel Schulte, Nadia Pastrone, Ken Long, CERN Cour. 60 (2020) 3, 41-46.

\bibitem{Mario16} Mario Greco, Tao Han, Zhen Liu, Physics Letters B  {\bf 763} (2016) 409-415.
\bibitem{Antonio20} Antonio Costantini, et al., J. High Energ. Phys.  {\bf 2020}, 80 (2020).
\bibitem{Dario18} Dario Buttazzo, et al., J. High Energ. Phys.  {\bf 11}, 144 (2018).
\bibitem{Nazar20} Nazar Bartosik, et al., arXiv:2001.04431.

\bibitem{muone1963} G. Backenstoss, et al., Phys. Rev.  {\bf 129}, 2759 (1963); T. Kirk and S. Neddermeyer, Phys. Rev. {\bf 171}, 1412 (1968); P. L. Jain and N. J. Wixon,  Phys. Rev. Lett. {\bf 23}, 715(1969).

\bibitem{muone1990} D. Adams, et al., Nucl. Instrum. Meth. A {\bf 443}, 1 (2000).

\bibitem{muone} G. Abbiendi, et al.,  Eur. Phys. J. C {\bf 77} (2017) 3, 139.

\bibitem{hou96} George Wei-Shu Hou, Nucl. Phys. Proc. Suppl. {\bf 51A} (1996) 40.  
\bibitem{choi97} Choi, S.Y, Kim, C.S., Kwon, Y.J., Lee, S.H., Phys. Rev. D {\bf 57} (1998) 7023.
\bibitem{barger97} V. Barger, et al., Phys. Lett. B {\bf 415} (1997) 200.

\bibitem{fabio20} F. Bossi,  P. Ciafaloni. Lepton Flavor Violation at muon-electron colliders. J. High Energ. Phys. 2020, 33 (2020).

\bibitem{CMS16} V.Khachatryan, et al., (The CMS Collaboration), Phys. Lett. B {\bf 763} (2016) 472.

\bibitem{ATLAS19} G. Aad, et al., (The ATLAS Collaboration), Phys. Lett. B {\bf 801} (2020) 135148.

\bibitem{BORONAT} M. Boronat, J. Fuster, et al., Phys. Lett. B, {\bf 750}(2015)95-99.

\bibitem{Qin17} Qin Qin, et al., Eur. Phys. J. C (2018) {\bf 78}: 835.

\bibitem{Alwall:2014hca} J.~Alwall {et al.}, JHEP {\bf 1407}, 079 (2014).

\bibitem{isr17} Qiang Li, Qi-Shu Yan, arXiv:1804.00125; Cheng Chen, et al., J. Phys. G {\bf 45} (2018) 1, 015004.

\bibitem{CLIChiggs} H. Abramowicz et al., Eur. Phys. J. C 77 (2017) 475.

\bibitem{Sjostrand:2003wg} T.~Sjostrand, L.~Lonnblad, S.~Mrenna and P.~Z.~Skands, hep-ph/0308153;    Comput. Phys. Commun. {\bf 191} (2015) 159.

\bibitem{deFavereau:2013fsa} J.~de Favereau {et al.}  [DELPHES 3 Collaboration],
  JHEP {\bf 1402}, 057 (2014).

\bibitem{shiltsev17} V. Shiltsev, arXiv:1705.02011.


\end{thebibliography}

\end{document}